# Elucidating the barriers on direct water splitting: Key role of oxygen vacancy density and coordination over PbTiO$_3$ and TiO$_2$

Ersen Mete[1,*], Şinasi Ellialtıoğlu[2], Oguz Gulseren[3], Deniz Uner[4,#]

[1] Department of Physics, Balıkesir University, Balıkesir 10145, Turkey
[2] Basic Sciences, TED University, Kolej, Ankara 06420, Turkey
[3] Department of Physics, Bilkent University, Ankara 06800, Turkey
[4] Department of Chemical Engineering, Middle East Technical University,
    Ankara 06800, Turkey

**In this work, using the state-of-the-art first principles calculations based on density functional theory, we found that the concentration as well as coordination of surface oxygen vacancies with respect to each other were critical for direct water-splitting reaction on the (001) surfaces of PbTiO$_3$ and TiO$_2$. For the water-splitting reaction to happen on TiO$_2$-terminated surfaces, it is necessary to have two neighboring O-vacancies acting as active sites that host two adsorbing water molecules. However, eventual dissociation of O–H bonds is possible only in the presence of an additional nearest-neighbor O-vacancy. Unfortunately, this necessary third vacancy inhibits the formation of molecular hydrogen by trapping the dissociated H atoms over TiO$_2$-teminated surfaces. Formation of up to 3 O-vacancies, is energetically less costly on both terminations of PbTiO$_3$ (001) surfaces compared with that of TiO$_2$, the presence of Pb leads to weaker O bonds over these surfaces. Molecular hydrogen formation is more favorable over the PbO-terminated surface of PbTiO$_3$, requiring only two neighboring oxygen vacancies. However, hydrogen molecule is retained near the surface by weak van der Waals forces. Our study indicates two barriers leading to low productivity of direct water splitting processes. First and foremost, there is an entropic barrier imposed by the requirement of at least two nearest-neighbor O-vacancies, sterically hindering the process. Furthermore, there are also enthalpic barriers of formation over TiO$_2$-terminated surfaces, or removal of H$_2$ molecules from the PbO-terminated surface.**

INTRODUCTION

Understanding and finding a route towards efficient water splitting reaction is not only important for energy and technological applications[1] but it is essential from fundamental point of view. Several different model systems like titania or perovskite surfaces have been proposed and have been investigated for this purpose.[2–4] Existing routes for hydrogen production from renewable energy include (i) using renewable electricity for electrolysis of water;[5] (ii) using concentrated solar energy for steam and/or carbon dioxide reforming of methane;[6,7] (iii) photo assisted catalytic or electrochemical water splitting.[8] If proven successful,

photo(electro)chemical production of hydrogen is the most attractive, due to the simplicity of the process and the potentials this simple process offers.[9]

Both direct photo-electrochemical and photocatalytic water splitting suffer from two problems. The first problem is the low production rates and photon conversion efficiencies. The second problem, which is also directly related to the first problem, is the difficulty with which the structures release oxygen for a complete catalytic cycle. In photo-electrochemical processes the bottleneck is the oxygen release. The thermodynamic driving forces under ambient conditions are towards water, hence once hydrogen and oxygen molecules are produced, their contact must be avoided in order to inhibit spontaneous combustion. This condition is present in electrochemical cells where anode and cathode are connected through a circuit and an electrolyte, but produced hydrogen and oxygen are collected separately. Still, in the electrochemical systems, the bottleneck is the oxygen evolution reaction.[10] The current consensus for the prevailing mechanism of oxygen evolution in oxides or perovskites is through the formation of oxygen vacancy. The subsequent steps of this process depend on several factors including the nature of the oxide material.

On the contrary, photocatalytic hydrogen production is carried out in slurry type of suspended solid catalysts in solvents. Well-mixed conditions, needed for an efficient reaction, also provides the environment for the immediate back oxidation of hydrogen to water. These processes can be inhibited by a careful design of the photo-reactor; or designing catalysts where the oxygen evolution and hydrogen evolution are delegated to different facets of the same crystallite.[9] Despite the efforts, hydrogen production rates are within a few milimol/h·g catalyst as revealed by a thorough review of literature, accompanied by the machine learning approaches.[11]

The dilemma in all these processes reside in the thermochemical bulk properties of the semiconductor oxides. The inherent photocatalytic activity of the semiconductor oxides depends on the capacity of bond breaking by the interaction of the oxide with electromagnetic radiation. The absorption of the electromagnetic radiation with the proper wavelength result in the creation of $e^-$–$h^+$ pairs. The chemical nature of these $e^-$–$h^+$ pairs depends on the oxide itself. For $TiO_2$ for example, upon irradiation, $e^-$ finds a place on Ti atom changing the oxidation state from +4 to +3, while the bond between the oxygen and Ti atoms are severed. In other words, the "softness" of the bonds between oxygen and the metal atoms is required for photon absorption, on the other hand, the softness is the reason that the formed hydrogen can interact with these bonds for oxidation. The chemical potential of hydrogen formed is sufficient to reduce the lattice oxygen.

One final barrier in the efficient production of hydrogen is to remove hydrogen formed by water splitting that is accommodated on the solid. Hydrogen can be accommodated in the form of spilled over hydrogen, that can be held around the surface or in the bulk.[12] The limitations can also be due to kinetic metastabilities dictated by a low surface coverage of hydrogen atoms. The spatial separation may inhibit formation of dihydrogen for desorption, hence lower the productivity.

The objective of this study is to search the limits of water-splitting reaction over $TiO_2$ and $PbTiO_3$ through first principles methods. We determined the barriers on the formation and removal of molecular hydrogen on titania and lead titanate. We started our analysis by assuming that oxygen vacancies are available, acting as two neighboring active sites hosting two water molecules. We examined the possibility of water dissociation depending on the coordination and clustering of surface O-vacancies by using *ab initio* methods.

**METHOD**

Density Functional Theory (DFT) calculations were performed using the VASP[13] within the framework of projector augmented-wave (PAW) method[14] based on plane-wave expansion of single particle states up to an energy cut-off value of 400 eV. The exchange and correlation effects were taken into account in the calculations by using the SCAN (strongly-constrained and appropriately-normed)[15] meta-GGA (semi-local meta-Generalized Gradient Approximation) density functional. SCAN functional lifts the over-binding in oxide materials which is common to GGA functionals.[16] The shortcoming of semi-local XC functionals in description of the electronic structures of materials associated with the self-interaction (SI) can be overcome by either admixing Hartree–Fock exchange with local density approximation (LDA) based exchange terms or supplemental on-site Coulomb interaction for localized electrons, a method known as LDA+$U$. In particular, the transition metal oxides with partially occupied $d$-states were erroneously predicted to be metallic by semi-local XC functionals. A proper description of both defect related gap states and the band gap can be obtained by imposing a suitably chosen $U$-parameter for Ti $3d$ states. Recently, use of $U$-parametrization for O $2p$ states in oxides was shown to enhance energetics and forces as to match those of hybrid functionals.[17] A systematic trace of $U$-parameter space showed that defect related electronic properties can be properly described if the SCAN functional is supplemented with parameters of $U = 6$ for Ti $3d$ and $U = 3$ for O $2p$ states. Calculated electronic band gaps of pristine $PbTiO_3$ and $TiO_2$ are 2.53 eV (exp. 2.5 eV[18]) and 2.77 eV (exp. 3.1 eV[18]), respectively. A 4×8 ×1 $k$-point mesh was used for Brillouin zone integrations. Atomic coordinates were fully relaxed by imposing a convergence criterion of $10^{-2}$ eV/atom for the minimization of Hellmann–Feynman forces on each atom in every spatial cartesian direction.

Both PbO- and $TiO_2$-terminations of $PbTiO_3$(001)-(3×2) and $TiO_2$(001)-(3×2) surfaces were modeled as slabs which were constructed from their bulk phases including 11 atomic layers along [001] direction to ensure the convergence of surface energies as suggested previously.[19] A vacuum region with a thickness of 12 Å along surface normal was introduced in the computational supercells to avoid any unphysical interaction between the periodic images of the slabs. Full geometry optimizations were performed without freezing any of the atoms to their bulk positions and without imposing symmetry. In addition, the post-processing of the calculated data was needed to determine the charge accumulation around each atom which was carried out using the Bader algorithm.[20]

**RESULTS AND DISCUSSION**

There are basically two possible terminations for the lowest index (001) plane of $PbTiO_3$. These are the PbO and $TiO_2$ surface layers. The latter has the lowest surface energy among all possible cleavages including higher-index planes.[21] Therefore, the (001) surface with $TiO_2$-termination is the most stable surface of $PbTiO_3$. When there are oxygen deficiencies on the surface, $TiO_2$-termination energetically becomes even more favorable than the PbO-termination. [22]

Water-splitting reactions, whether thermochemical, electrochemical, or photocatalytic, involve reduction of oxide materials through an energy source such as heat,[23,24] electrons, or photons. Therefore, oxygen vacancies play an important role in capturing and splitting water molecules.[25–28] We explored the necessary requirements for direct splitting of water molecules via surface oxygen vacancies on $PbTiO_3$ and $TiO_2$. The conditions depend not only on the material but also on the termination due to different surface chemical reactivities.

$TiO_2$-terminated top layer of $PbTiO_3$ mimics titania surfaces and provides a setting for a variety of chemical reactions. For instance, molecular adsorbates tend to exhibit similar adsorption characteristics as on the anatase (001) surface.[29,30] Bonds usually form between their tail oxygens and the surface Ti atoms. This looks like a continuation of bulk-like $TiO_2$ structure. In an experimental realization, the surface might have oxygen vacancies which leave excess charge around the nearest-neighbor Ti atoms. Such vacancy sites become more active in capturing molecules leading to relatively stronger binding.

The energy required to remove an oxygen atom from the surface is 3.86 eV on the PbO-termination and 3.96 eV on the $TiO_2$-termination. It is energetically 0.1 eV more favorable on the PbO-terminated surface. The formation energy of an oxygen vacancy on the surface of titania is 4.39 eV (Table 1). We explored necessary conditions for hydrogen molecule formation as a result of oxygen vacancy-induced splitting of water on the surfaces. In order to examine the possibility of splitting of two adjacent water-molecule adsorbates to their hydrogen atoms and hydroxyl groups on the surfaces, we considered a number of configurations. As a starting point, the minimum condition appears to be the availability of the adjacent surface oxygen vacancies which act as an adsorption center. The adsorption of two water molecules at the two adjacent oxygen vacancy sites allows a close proximity for the nearest-neighbor hydrogen atoms which face each other at the lateral direction. The energy cost of formation of two and three neighboring surface oxygen vacancies, in Table 1, increases depending on the material and the surface-termination, the formation energy per vacancy appears to slightly fluctuate around the value of single vacancy.

**Table 1.** Formation energies (in eV) of nearest-neighbor surface oxygen vacancies calculated using SCAN XC functional.

| Surface Model | 1 $O_{vac}$ | 2 $O_{vac}$ | 3 $O_{vac}$ |
|---|---|---|---|
| $TiO_2$(001)-(3×2) | 4.39 | 9.77 | 15.39 |
| $PbTiO_3$(001)-(3×2) $TiO_2$-terminated | 3.96 | 7.39 | 11.49 |
| $PbTiO_3$(001)-(3×2) PbO-terminated | 3.86 | 8.12 | 11.37 |

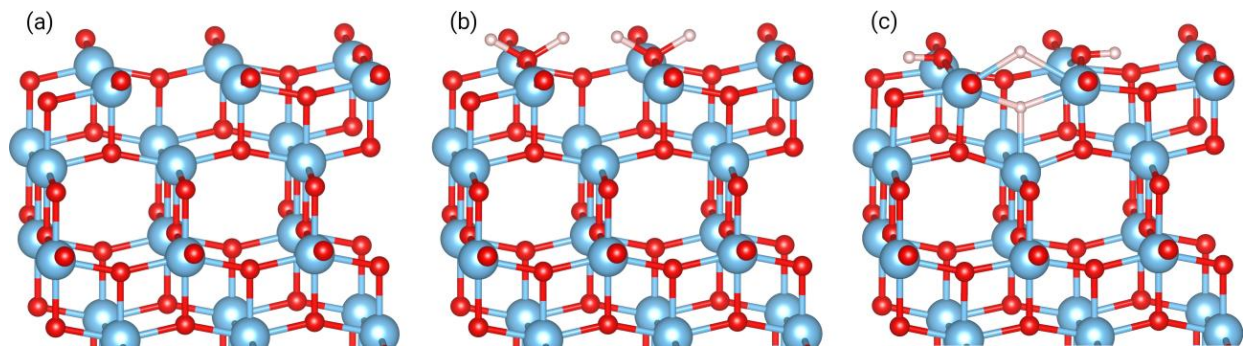

Figure 1. Schematics of TiO$_2$(001)-(3×2) (a) with a cluster of 3 adjacent oxygen vacancies (O$_{vac}$) only; (b), (c) with additional 2 H$_2$O adsorbates at the neighboring oxygen vacancy sites in their corresponding initial and final geometries, respectively. Light blue (Ti), red (O) and white (H) colors are used for the ball-and-stick illustration of the model structure.

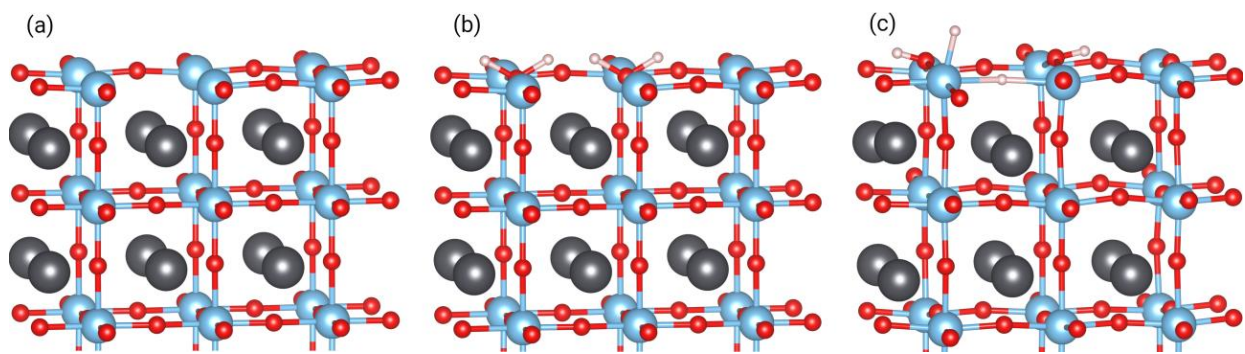

Figure 2. Schematics of PbTiO$_3$(001)-(3×2) TiO$_2$-terminated (a) with 3 adjacent oxygen vacancies (O$_{vac}$) only; (b), (c) with additional 2 H$_2$O adsorbates at the neighboring oxygen vacancy sites in their corresponding initial and final geometries, respectively. Dark gray (Pb), light blue (Ti), red (O) and white (H) colors are used for the ball-and-stick illustration of the model structure.

In the cases of TiO$_2$-terminations on both titania and lead titanate, the adsorption geometry of two water molecules at the two neighboring oxygen vacancy sites is shown in Figure 1(b) and Figure 2(b), respectively. Formation of Ti–O bonds as a result of charge transfer from the surface Ti to oxygen atom due to water adsorption, leads to a weakening of O–H bonds within the water molecule. Bader analysis, as a comparison of the total charge on the substrate before and after the adsorption, show a charge transfer of 1.43 and 1.39 electrons from the titania and TiO$_2$-terminated lead titanate slabs, respectively. This type of surface oxidation and atomistic configuration solely is not enough to achieve water splitting. The DFT calculations indicated that there is a need for an additional oxygen vacancy which must also be adjacent to both of the adsorption sites as shown in Figure 1(b) and Figure 2(b). Adsorption of water on the surface loosens the O–H bonds. Moreover, the water molecules slightly lean toward each other due to absence of lateral Ti–O bonds between them. Then, the distance between the H atoms facing each other becomes as small as 1.8 Å in the lateral direction due to the presence of the third oxygen vacancy. This site has to remain as a vacancy in order to ensure breaking the O–H bonds. This mechanistic picture gives rise to the dissociation of H atoms from their parent water molecules leaving the hydroxyl groups behind. The ratio of the clustering vacancies to the number of surface oxygens is 1/3. When we considered a cluster of four neighboring surface oxygen vacancies, the geometry optimization ended up a significant local distortion which drives the reduced surface layer to undergo a reconstruction. These kinds of defect driven structural changes due to oxygen vacancies are known for the formation of black $TiO_2$.[31,32] Although, the H atoms which were dissociated from their parent water molecules are close to each other, they did not form an $H_2$ molecule. On the $TiO_2$-termination, those hydrogen atoms were, however, attracted to the adjacent oxygen vacancy sites forming different adsorption geometries for titania and lead titanate due to differences in their chemical environment at the sub-surface layers as shown in Figure 1(c) and Figure 2(c). One of the main differences is the undulation geometry of $TiO_2$ network of the top layer which mechanistically plays an important role in the subsequent pathway of the dissociated H atoms. The Ti–Ti separations on the surface layer are 3.79 Å and 3.89 Å on $TiO_2$(001) and $PbTiO_3$(001), respectively. Energetically, the adsorption of a water molecule at a surface oxygen vacancy site is considerably stronger on titania. The calculated adsorption energies are –1.43 eV and –1.11 eV on $TiO_2$(001) and $TiO_2$-terminated $PbTiO_3$(001). On the surface of titania, both of the H atoms were trapped at the vacancy site reflecting a Ti–H bonding character as shown in Figure 1(c). Bader charge analysis show an accumulation of 1.66 electrons around this H atom indicating a partial filling of the 2$s$ orbital. In this geometry, the position of lower lying H mimics surface-like oxygen coordination with the nearest-neighbor Ti atoms. On the surface of lead titanate, one of the H atoms is attracted to the oxygen vacancy while the other H atom is weakly bound atop the nearest-neighbor surface Ti atom (Figure 2(c)). Bader charges are 1.66 and 1.58 electrons for these H atoms, respectively. Both H atoms were trapped on the surface layer preventing formation of $H_2$. Then, breaking of the translational symmetry induced by these impurities on the surface layer influences the sublayer such that Pb atoms are distorted from their bulk-like centroidal positions.

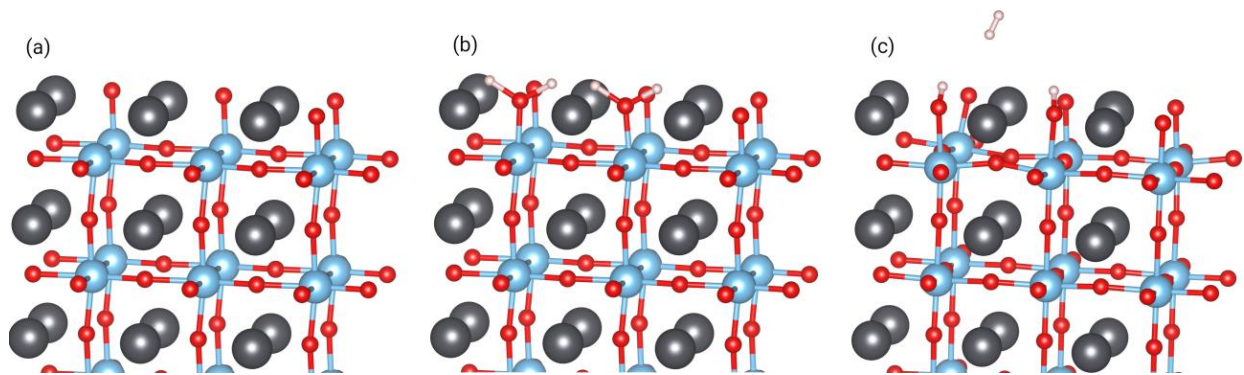

Figure 3. Schematics of $PbTiO_3$(001)-(3×2) PbO-terminated (a) with 2 adjacent oxygen vacancies ($O_{vac}$) only; (b), (c) with additional 2 $H_2O$ adsorbates at the neighboring oxygen vacancy sites in their corresponding initial and final geometries, respectively.

In the case of PbO-terminated surface of PbTiO$_3$, when two water molecules are adsorbed on two nearest-neighbor oxygen vacancy sites as shown in Figure 3(b), they slightly lean toward each other and the O–H bonds loosen as a result of charge transfer from the substrate to the adsorbate Bader analysis gave a value of 0.85 electrons for lead titanate. Then, the H–H distance becomes 1.8 Å similar to the previous cases. The adsorption energy per water molecule at a surface oxygen vacancy site is 0.97 eV which is significantly lower than those of TiO$_2$-terminations. Another important difference of PbO-termination from the TiO$_2$-terminations is the coordination in the corresponding surface layers. The Pb–O bonds are 2.54 Å on the PbO-termination while Ti–O bonds are 2.02 Å on the TiO$_2$-termination. This indicates a stronger coordination in the TiO$_2$ network. These factors can be attributed to the dissociation of H atoms from their parent water molecules because of the attraction between the two hydrogen atoms in close proximity to each other. They form an H$_2$ molecule leaving the hydroxyl groups at the vacancy sites, because the required charge compensation is supplied by the Ti atoms which were exposed at the adsorption site. Bader charge analysis showed that the charge on each H atom which contributes to the covalent bonding in the hydrogen molecule is one electron while the charge on the H atom of the hydroxyl group is zero. Then, the hydrogen molecule stays above the surface plane at a van der Waals distance (2.90 Å) due to absence of an additional surface oxygen vacancy near adsorption site, as shown in Figure 3(c).

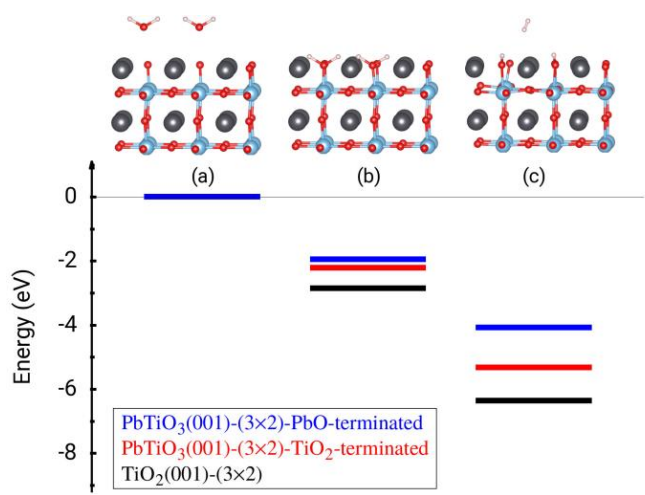

Figure 4. Energy level diagram of three stages of water splitting on PbTiO$_3$ and TiO$_2$ surfaces: (a) surface with adjacent oxygen vacancies and water molecules away from the surface, (b) water molecules get adsorbed, (c) water splitting. Ball-and-stick illustration shown for PbO-terminated surface of PbTiO$_3$ as a representative.

The energy level diagram for the pathway of water adsorption and dissociation on the oxygen deficient surfaces is shown in Figure 4. Consistent with the calculated adsorption energies, first stage from (a) to (b) describes separate adsorption of two water molecules on two nearest-neighbor oxygen vacancies. Binding of water on TiO$_2$ is significantly stronger, which indicates weakening of lateral Ti–O bonds associated with Pb atoms in the case of PbTiO$_3$. Pb suppresses the undulation of Ti–O layer. The lateral Ti–O bonds are 1.94 Å and 1.99 Å in titania and lead titanate, respectively. For the second stage from (b) to (c), $TiO_2$-terminated surfaces appear to be energetically more favorable for water splitting. We note that dissociation of water did not take place on $TiO_2$-terminations without an additional oxygen vacancy which connects the water adsorption sites. Therefore, at least a cluster of three oxygen vacancies is necessary on $TiO_2$-terminations. However, dissociation of water did not end up with $H_2$ production on the (001) surfaces of neither $TiO_2$ nor $TiO_2$-terminated $PbTiO_3$ for the oxygen vacancy mediated models considered here. The energy differences from (b) to (c) are, –3.12 eV (PbO-termination of $PbTiO_3$), –2.13 eV ($TiO_2$-termination of $PbTiO_3$), and –3.51 eV ($TiO_2$). The only case which can produce $H_2$ appears to be the PbO-termination with two adjacent oxygen vacancies which were occupied with two separate water molecules even though the energy difference in the last stage is smaller than the other cases.

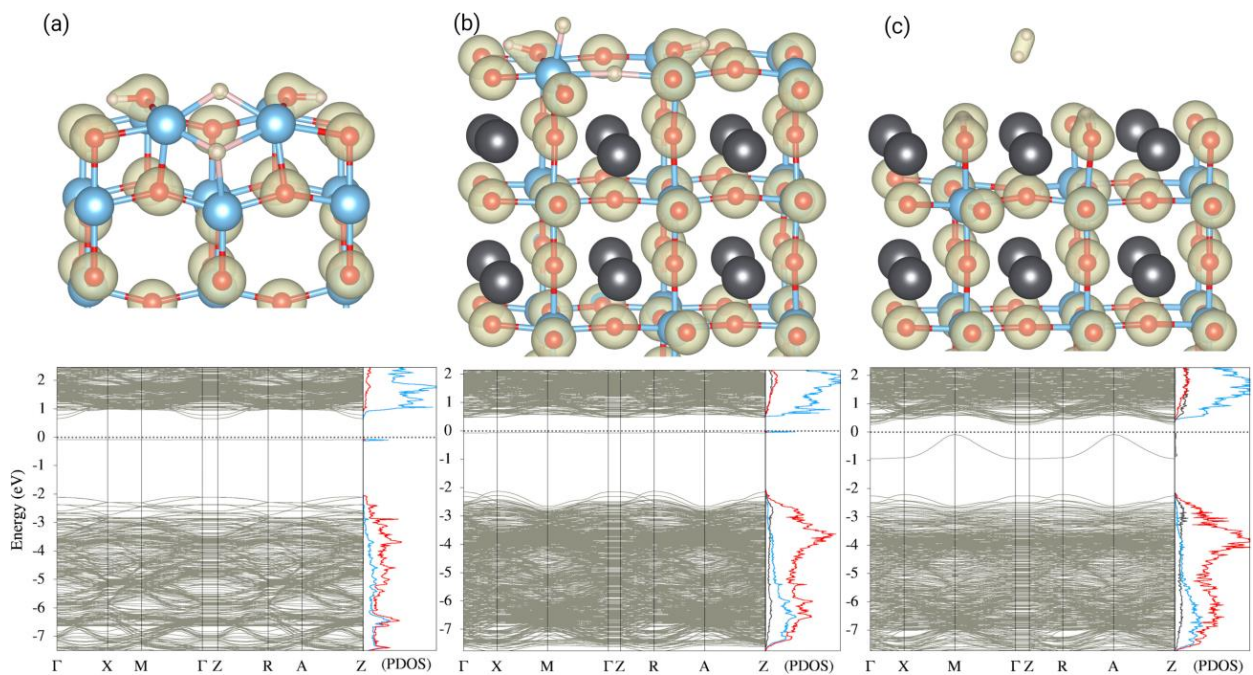

Figure 5. Charge density plots and corresponding electronic band structures with partial densities of states (PDOS) of (a) $TiO_2(001)$-(3×2) + 3 $O_{vac}$ + 2 $H_2O$ system, (b) $PbTiO_3(001)$-(3×2)-$TiO_2$-terminated + 3 $O_{vac}$ + 2 $H_2O$ system, (c) $PbTiO_3(001)$-(3×2)-PbO-terminated + 2 $O_{vac}$ + 2 $H_2O$ system calculated using the SCAN+$U$ ($U$ = 6 for Ti $3d$ and $U$ = 3 for O $2p$) approach. A color scheme of dark gray, light blue, red and white is used for Pb, Ti, O, and H, respectively.

Total charge distributions, the energy bands, and the corresponding partial densities of states (PDOS) were calculated as shown in Figure 5. Charge density plots show the bonding characteristics of $TiO_2$ and $PbTiO_3$ lattice structures. Calculated Bader charges are in consistency with the nominal charge states for the lattice cations and anions as $Ti^{4+}$, $O^{2-}$, $Pb^{1+}$. More importantly, the charge distribution after water dissociation reveals the bonding of dissociated H atoms to either to the substrate or to each other to form $H_2$. Especially, the charge around the H atoms, which were bound to surface Ti atoms, indicate a partial filling of the $2s$ orbitals.

Energy bands and electronic densities of states were calculated using SCAN functional with supplementary Hubbard-$U$ term to better describe the repulsion between strongly correlated Ti $3d$ and, to a lesser degree, O $2p$ electrons. This is one of the methods necessary to overcome the band gap underestimation of the standard XC functionals within the framework of DFT. In addition, the nature of defect related states must be reasonably defined in the gap. The use of SCAN functional instead of standard GGA functionals was not enough to get the band gaps of $TiO_2$ and $PbTiO_3$ properly. Similar findings were reported previously by a theoretical study.[17] The choice of supplementary empirical $U$ term becomes critical and needs to be validated for these oxide materials in the presence of point defects. After testing several $U$ parameters for Ti $3d$ with and without O $2p$ orbitals, the optimal choice was found as $U = 6$ for Ti $3d$ and $U = 3$ for O $2p$ states, which not only heals the band gap but also gives a sound description of the defect related gap states for both of the materials (please see Supplementary Information). In particular, the application of Hubbard-$U$ for O $2p$ orbitals become important for the electronic description of dissociated H atoms on a cluster of three oxygen vacancies on the $TiO_2$-termination of $PbTiO_3$.

The valence band (VB) dominantly reflects Ti–O bonding character with a contribution from Pb–O bonds in the case of $PbTiO_3$. The upper part of VB is characterized by O $2p$-electrons, which is common and well known for transition metal oxides. The conduction band (CB), on the other hand, dominantly contributed by anti-bonding Ti $3d$ states. In the cases of $TiO_2$-terminated surfaces shown in Figure 5(a) and 5(b), two hydroxyls are bound to two adjacent vacancies. This binding appears as a doubly degenerate state at the top of VB reflecting Ti–O bonding character. In addition, trapping of dissociated H atoms at the third oxygen vacancy brings a flat going occupied state 0.73 eV and 0.55 eV below the conduction band minimum (CBM) of $TiO_2$ and of $PbTiO_3$, respectively. Well localization of these defect related states indicates a weak binding of dissociated H atoms to the vacancy site on the substrate. In the case of PbO-termination of $PbTiO_3$ shown in Figure 5(c), the valence band maximum (VBM) is characterized by the hydroxyl–vacancy bond similar to the previous structures. A gap state 0.34 eV [Γ–M] below the CBM appears due to $H_2$. The dispersion of this state along $k_z$ is consistent with the H–H covalent bond which lies in the $z$-direction (or [001]) with respect to the surface plane. The SCAN+$U$ method successfully explains and gives insights about the origin and nature of electronic energy band structure of the models which were constructed to represent water splitting on $TiO_2$ and $PbTiO_3$.

**CONCLUSIONS**

The number and clustering of surface oxygen vacancies becomes critically important for direct water splitting on the (001) surfaces of $TiO_2$ and $PbTiO_3$. Presence of Pb weakens binding of lattice oxygens. Therefore, the energy cost of oxygen vacancy formation on $TiO_2$ is higher than that on $PbTiO_3$. The minimum requirement appears to be the adsorption of two water molecules on two nearest-neighbor oxygen vacancy sites separately, for the possibility of a subsequent dissociation of their H atoms which face each other. $TiO_2$-terminated surfaces need a cluster of three oxygen vacancies for which two of them act as water capturing active sites and the remaining one must be adjacent to both of them and must remain unoccupied. The existence of this third vacancy is necessary for dissociation of the H atoms from their parent water molecules yet is detrimental for the formation of $H_2$ since it acts as a trap for dissociated H atoms. Revelation of this bottleneck is useful in explaining the production of very low quantities of $H_2$ by direct water splitting on $TiO_2$-terminations which are otherwise known for their excellent catalytic properties. On the PbO-termination of $PbTiO_3$, two adjacent oxygen vacancies can capture two water molecules which lean toward each other. Based on the DFT calculations vacancy-mediated water splitting seems to be possible on PbO-termination. However, hydrogen molecule must overcome the van der Waals attraction towards the surface.